# Prospects for an advanced Kennedy-Thorndike experiment in low Earth orbit

J. A. Lipa, S. Buchman, S. Saraf, J. Zhou, A. Alfauwaz, J. Conklin, G. D. Cutler and R. L. Byer,

Hansen Experimental Physics Laboratory, Stanford University, Stanford, CA.

We discuss the potential for a small space mission to perform an advanced Kennedy-Thorndike test of Special Relativity using the large and rapid velocity modulation available in low Earth orbit. An improvement factor of ~100 over present ground results is expected, with an additional factor of 10 possible using more advanced technology.

## INTRODUCTION

It is well known[1] that Special Relativity (SR) owes its experimental foundation to a minimum of three types of classical experiments: those of Michelson and Morley[2] (MM), Kennedy and Thorndike[3] (KT) and Ives and Stilwell[4]. Extreme precision has been obtained in verifying SR in many areas[5], yet the basis for accepting it must remain experimental, as with all laws of physics. Recently a new approach has been developed[6,7] to categorize potential manifestations of Lorentz invariance violations (LIV) that might occur within the Standard Model (SM). A generalized extension to the SM (SME) was developed and a classification scheme for possible small effects was presented. With the restriction to the photon sector of the SM and renormalizable terms, a sub-group of the coefficients of LIV was found to be related to the putative angle dependence of speed of light ($c$) within a Sun-centered near-inertial reference frame[7]. This implied their detectability with MM type experiments. With the same restrictions, an analysis of KT type experiments yielded similar terms. However from a broader perspective it is now recognized[8] that higher order terms are critical to the analysis of these latter experiments. An essential feature of a KT experiment is to study the boost-dependence of $c$ rather than its angle dependence.

To date, efforts to place limits on LIV related to KT effects have relied on the Earth's rotational and orbital motions to modulate the velocity of an apparatus relative to some inertial reference[9]. The resulting limits on the constancy of $c$ are on the order of $\delta c/c <\sim 10^{-15}$. While improvements in technology will no doubt lead to further gains, it has become clear that space experiments in low Earth orbit offer a way to obtain much better results than with ground experiments. In the remainder of this paper we discuss the design and limiting factors of such an experiment involving a small free-flying satellite in low Earth orbit. We conclude that a conservative factor of ~ 100 gain is possible in a 2-year mission. A related mission, OPTIS, was proposed some time ago to fly in a higher altitude elliptical orbit[10]. An earlier form of the present mission has been described[11]; here we update the mission design and emphasize the scientific and technical underpinnings.

## BACKGROUND

Among other assumptions, Einstein's SR posits that $c$ is exactly constant, rigorously independent of the magnitude and direction of the velocity of the observer relative to any rest frame and independent of observation angle in the local frame. These invariance features of $c$ are used to tie space and time together, leading to the famous Lorentz transformations. These ideas work remarkably well, but difficulties appear when one tries to combine Einstein's theories with the SM. Since some fundamental theories attempting to bridge this divide (such as quantum gravity) allow large violations of LI at high energies such as those encountered in the Big Bang era, a natural question is: can remnants of these effects be detected in today's universe? If detected, a small LIV would point to new phenomena beyond the SM of particle physics and provide evidence for a preferred direction or reference frame in the cosmos. A new fundamental understanding of space-time and matter would result, and our view of many astrophysical phenomena could be altered. Dramatic advances are likely in areas directly connected to this physics: behavior of matter and radiation in extreme physical conditions such as the Big Bang, very near black holes, and in the early Universe – as informed by observations of electromagnetic and gravitational radiation. Cosmology could be revolutionized if, for example, the frame in which the Cosmic Microwave Background (CMB) is isotropic were found to be preferred in some deeper way.

The intrinsic boost-dependence of a KT experiment can easily be seen from the idealized model of Mansouri and Sexl[12] (MS) that separates small LIV terms into angle and boost dependent groups. Analyzed by Robertson[1] and MS, the RMS model of Lorentz violations can be parameterized as:

$$\delta c/c = \varepsilon_{KT}(v/c)^2 + \varepsilon_{MM}(v/c)^2\sin^2\theta \qquad (1)$$

where the quantities $\varepsilon_{KT}$ and $\varepsilon_{MM}$ are small dimensionless coefficients, $\theta$ is the angle of propagation of light relative to some preferred direction, $v$ is the velocity of the apparatus relative to the frame, and $\delta c$ is the deviation of $c$ from an exact constant. In SR the two terms on the right hand side are zero. In the RMS model a MM experiment can be viewed as measuring the quantity $\varepsilon_{MM}$ via the $\theta$-dependence of $c$, while a KT experiment measures the quantity $\varepsilon_{KT}$ via the velocity dependence of $c$, independent of $\theta$. A generalized experiment may be sensitive to both terms.

A weakness of the RMS model is that it ignores the physical processes occurring in real clocks and rods used to make the measurements. The modern approach to LIV addresses this problem and is based on cataloging all possible sources of violation via the SME. This work has produced many new insights to existing experiments and has been shown to include the RMS model. The number of parameters has been greatly expanded to include effects from the entire fermion and boson sectors of the SM. With the restrictions mentioned in the introduction Kostelecky and Mewes[7] (KM) showed that $\varepsilon_{KT}$ and $\varepsilon_{MM}$ are given by similar sets of LIV coefficients. However, these restrictions overly constrain the measurement possibilities. The full set of possibilities in the SME includes multiple infinities of terms, many of which have not yet been cataloged. A more general interpretation of the experimental situation has been given[8] which indicates that non-trivial extensions to other sectors of the full SME are required. A more complete catalog of the SME terms involved is being developed. Until this situation is clarified, it appears best to interpret at least KT-style measurements within the RMS formalism, while acknowledging that a detailed SME interpretation will be forthcoming. Since other sectors of the SME are known to be involved, it is important to always give full details of the rods and clocks used[8].

It has been suggested[6] that the appropriate scale for remnants of Big Bang physics could be near the ratio of the electroweak mass to the Planck mass (quantities in the SM), $\sim 2 \times 10^{-17}$, implying that there could be modulations of the velocity of light $\delta c/c$ at this level. While not a firm prediction, this consideration adds interest to the measurements with the possibility of exciting new physics related to the SM. An LI violation can be viewed as a symmetry breaking effect and would undoubtedly transform physics on all energy scales, while an improved null result could constrain theories attempting to unite particle physics and gravity. If a violation were discovered then modifications to the SM, SR and General Relativity would be required. Since theoretical considerations point to a violation being more important at the high energies characteristic of the Big Bang, the impact on our view of the birth and evolution of the universe could be dramatic. Even a null result is of value because it puts tighter limits on the size of a violation that can be allowed in new theories attempting to unify the SM and General Relativity.

Recent ground experiments[9] have obtained the result $\varepsilon_{KT} = -4.8 \pm 3.7 \times 10^{-8}$ by comparing the frequencies of a cryogenic sapphire oscillator with a hydrogen maser over a period of about 6 years. In other areas of LIV, the Fermi Gamma Ray Space Telescope has the ability to time the arrival of gamma-ray burst signals with an uncertainty of $\sim 1$ millisecond, allowing the detection of one possible form of violation, namely a slight dependence of the speed of light on photon energy. Such photon dispersion might possibly arise from similar physical principles as anisotropy, but is very different from the observables that we will measure. Some interesting limits have already been obtained[13] on the energy dependence of $c$. Other limits can be obtained from observations of the isotropy of the flux of very high-energy cosmic rays[14]. The theoretical aspects of such experiments have been discussed by Coleman and Glashow[15]. No new limits for $\varepsilon_{KT}$ have yet been derived from these experiments. A useful summary of a broad range of earlier measurements bearing on LIV was given by Mattingly[16].

The mission described here will search for variations in $c$ at the level $\delta c/c < 10^{-17}$ over the entire sky with the possibility of an order of magnitude higher resolution depending mostly on optical cavity performance. It will be able to make a 100x improvement of the limit on the KT coefficient in eq. 1 and a 15x improvement on some of the parameters in the SME. The MM coefficient and other SME parameters will also be measured with resolutions at least as good as on the ground. The MM data also support the KT measurements by providing crosschecks and cover the slight possibility that measurements in space may be different to those on the ground for reasons as yet unknown[17]. The concept for the experiment is illustrated in figure 1. The KT measurements are enabled by the modulation of $v$ in eq. 1 by the orbital motion of the spacecraft and the precession of its orbit plane. The MM measurements are enabled by spinning the spacecraft at a low rate around the line of sight to the sun.

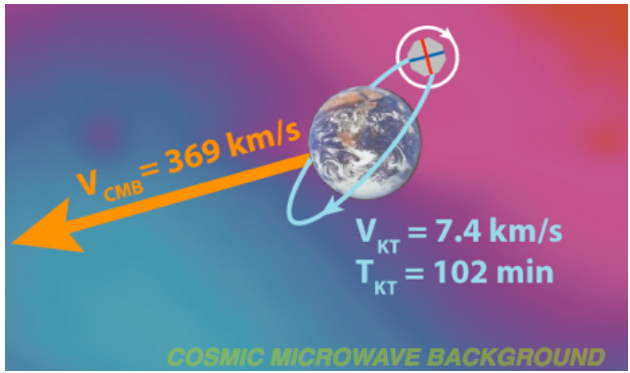

Fig. 1: Concept of STAR experiment.

Performing the experiment in space offers a number of advantages over ground experiments: With a well-designed spacecraft the vibration environment can be extremely quiet. Large dc gravity forces are eliminated, allowing the possibility of very soft mounting of force-sensitive components. Of course, latching during launch is necessary, but these can be retracted during science measurements. The wide range of orbits available allows much better optimization of the experimental conditions than an equivalent ground experiment. For a low earth orbit (LEO) the velocity modulation in eq. 1 is typically 16x greater than in a ground experiment using the Earth's rotation. The correspondingly short period of the velocity reversals is another major advantage because it greatly mitigates the thermal control problem. Also, it allows a large reduction in the secular drift requirements for the rods and clocks. A dawn/dusk sun-synchronous orbit has two further advantages: the spacecraft can be in sunlight for almost the entire year minimizing thermal effects, and orbit plane precession can be used to sweep the entire sky for velocity effects. The various motions involved also allow the separation of some systematic effects from signals aligned with inertial coordinates. On the other hand, since the spin axis is not aligned with gravity there are new perturbations that enter, primarily from gravity gradient and associated red shift. These are discussed in more detail below.

## EXPERIMENT DESIGN

The basic idea for the flight instrument is shown in figure 2. There are four input signals, each emanating from a common source—the 'clock'—a laser stabilized to hyperfine component $a_{10}$ in the $R(56)32$-$0$ transition at 532 nm of molecular iodine ($I_2$). This choice of transition allows us to use a laser that is already flight qualified. The first and fourth signals are the baselines used to compare the signals from the rods.

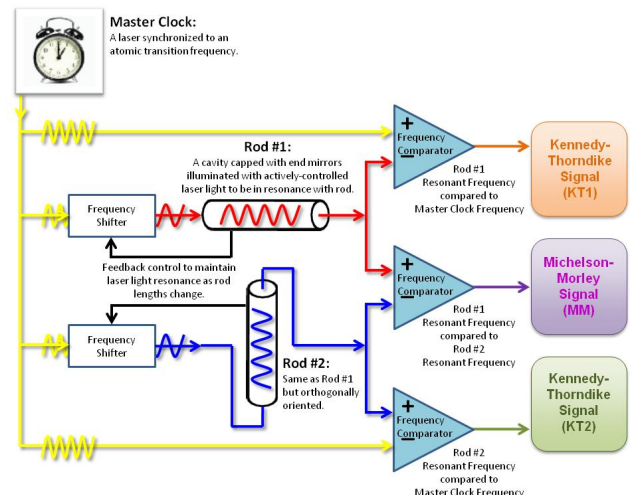

Figure 2: Conceptual diagram of instrumentation

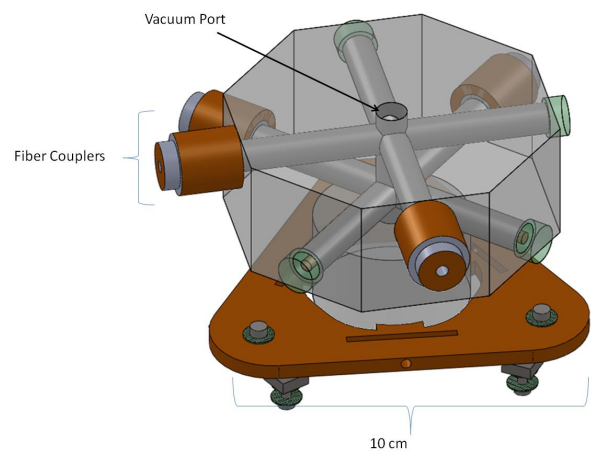

Figure 3: ULE optical cavity block with two sets of orthogonal fiber-coupled cavities (courtesy Ball Aerospace).

They go directly from the molecular clock to a pair of frequency comparators. The second signal is fed into a resonant cavity—'rod 1'—via a frequency shifter. The cavity and the shifter are in a feedback loop such that the frequency shifter can maintain the light frequency at the proper cavity resonance at all times. Either anomalous length contractions in rods 1 and 2 or time dilations will manifest themselves as deviations in the required frequency shift to keep the rods resonant with the laser light. The third signal is fed into a resonant cavity—'rod 2'—arranged orthogonally to rod 1. Like rod 1, rod 2's laser light is also maintained at resonance via its own frequency shifter. The outputs of the rod laser light are then compared with the baseline clock signals to generate two KT science outputs. Both rods are made of temperature-stable ultra-low expansion glass (ULE) to reduce spurious thermal effects.

The outputs of the KT frequency comparators are monitored as a function of the spacecraft's position around its orbit. The reversal of the spacecraft's velocity vector generates a sinusoidal 'boost' relative to inertial space on which any LIV signal should be imprinted. If there is no orbital variation of the recorded values, the KT portion of the experiment is null. If there is an orbital variation, the KT coefficient could then be measured. The MM science output is obtained by sending the signals from the two rods to another comparator. Its output is monitored as a function of orientation of the rods relative to inertial space. Since the spacecraft is slowly rotating, any signal related to MM would show up as a variation at approximately twice the spin rate. In the actual experiment, of course, all signals are recorded continuously and fit with sine and cosine series.

**Instrument requirements**

Resonant cavity interferometric methods represent the one of the few, if not only, means of detecting anisotropy in the velocity of light at the sensitivity levels required (i.e. $\delta c/c < 10^{-17}$ over a two year integration time). A pair of orthogonally oriented, very high finesse Fabry-Perot cavities serves as the sensors to detect such anisotropy. As the satellite rolls and orbits the Earth, asymmetries in space-time will manifest themselves as variations in the resonant frequencies of the cavities relative to each other or the clock.

The cavities are maintained at resonance using the Pound-Drever-Hall (PDH) locking technique[18]. The technique allows automatic adjustment to the frequency offset to keep the cavities at resonance in the presence of anisotropy. To meet the design goal of the experiment the length stability of the optical cavity has to be $\delta l/l < 5 \times 10^{-16}$ at orbit averaging time. In order to meet the mechanical stability requirements over an orbital period, the cavity assembly is encased in a thermally controlled environment of advanced design. The fiber-coupled cavities are nominally resonant at the 1064 nm laser wavelength that is stabilized to a hyperfine spectroscopic absorption line of Iodine. The laser frequency stability also needs to be $\delta f/f < 5 \times 10^{-16}$ at orbit averaging time. Finally, the frequency comparator accuracy needs to be $\delta f/f < 2 \times 10^{-16}$ at orbit time. These are the primary sources of random error in the measurement and are tabulated in table 2 along with other sources of error.

**Optical cavity block**

The effectiveness of the cavities as anisotropy sensors depends strongly on the line width and stability of the laser sources, the thermal-mechanical stability of the cavities themselves, and the sensitivity of the cavities to anisotropy-induced length perturbations. The last correlates most directly to cavity finesse, defined as the ratio of the free spectral range of the cavity to the full width at half maximum of the cavity resonance peak. The required finesse is $\sim 10^5$ to detect length perturbations at sensitivities necessary to meet the science objective. Modern cavities designed for laser locking[19] readily reach finesse values above $3 \times 10^5$ using pairs of low-absorption (<2 ppm), partially transmitting (~10 ppm) mirrors.

The cavities are made of temperature-stable ULE glass to minimize temperature-induced length fluctuations. The thermal behavior of this glass is the primary driver for the material selection of the cavity body[20]. ULE has a low coefficient of thermal expansion (CTE) of $\sim 10^{-8}$/K within an operating temperature range of 10 – 20 °C, and a null CTE near 15 °C. By operating the thermal enclosure within 3 mK of the CTE null it is possible to meet the $5 \times 10^{-16}$ length stability requirement at orbit period needed to achieve the science objective. A sketch of the planned cavity assembly is shown in figure 3.

**Thermal enclosure**

The thermal enclosure houses the optical cavity assembly and controls its thermal and structural environment. The most important requirement for this assembly is to filter the orbital and spin period temperature variations of the spacecraft to a level so low that the science signals from the optical cavities are unperturbed. Since the spin period is much shorter than the orbital period, this requirement mostly applies to the latter. The location of the CTE null of ULE is determined by its chemical composition. This allows the null to be tailored to near room temperature. Due to residual irregularities in production, the likely effective lower limit for an entire block of ULE is a CTE of ~ $10^{-9}$/K. For random temperature fluctuations, the constraint is $\delta l/l < 5 \times 10^{-16}$ at orbit averaging time, as mentioned earlier. However, when averaged over 2 years, we require $\delta l/l < 10^{-17}$. Thus it is clear that we need the cyclical long term systematic temperature variations of the cavity block to be $< 10^{-8}$ K at orbital period. This extreme requirement has driven us to design a 6-stage thermal isolation system with both active and passive control on various layers.

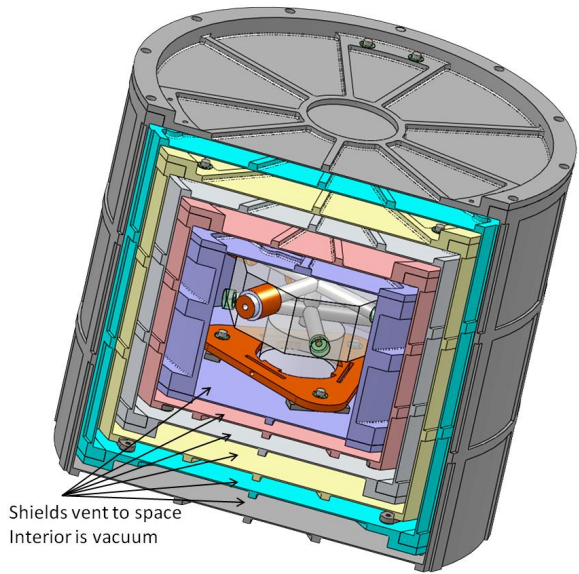

Fig. 4: Six-layer thermal enclosure cutaway showing the ULE optical cavity block.

A detailed thermal model of the enclosure and optical cavities based on the design in Figure 4 shows that a ±1 K temperature variation at the external interface will produce a temperature variation of less than ±1.5 nK at the optical block (see Fig. 5) without using active thermal control. While current analysis suggests that this passive thermal control system will exceed requirements, active thermal control has been included in the baseline design to provide additional margin and to allow control of the steady state temperature of the optical block to be near the CTE null.

A single thermal controller on an outer stage of the shielding can easily gain a further factor of 100 attenuation. Additionally, for the preferred near-polar orbit, spacecraft thermal modeling has shown that external temperature swings of <10 mK are expected at orbital period. We expect that this number is somewhat optimistic because the albedo's assumed were uniform around the earth. We have also modeled an equatorial orbit as a near-to-worst case for thermal perturbations. We find that with judicious design of thermal couplings it is possible to hold the temperature swing from light to shadow below 0.2 K at the surface of the thermal enclosure. Thus we show ample margin in the thermal design for essentially any orbit.

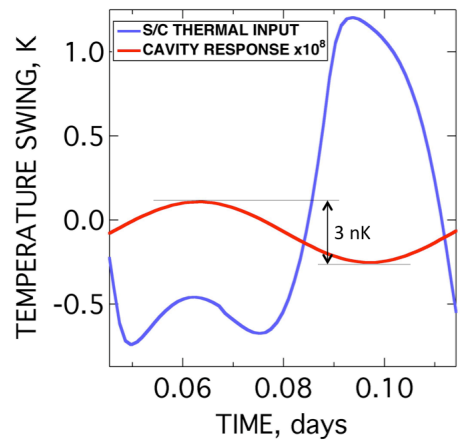

Fig. 5: Results from thermal modeling the 6-stage passive thermal enclosure in Fig. 4. The response of an optical cavity block (Fig. 3) over 1 orbit has a pp swing of 3 nK. The input signal (blue) was computed from a thermal model of the spacecraft in an equatorial orbit.

A structural analysis of the thermal enclosure and optical cavity block shows that the design has the adequate stiffness to survive launch, while greatly attenuating the external thermo-mechanical stresses placed on the optical cavity.

The design of the thermal enclosure in Figure 4 is based on a set of 6 nested Aluminum cans with Titanium alloy supports on alternating ends. The purpose of this arrangement is to greatly attenuate the stress effect of thermal expansion at the outside layer on the cavities. This attenuation factor is modeled to be $> 10^{15}$. This allows 10 K swings at the outer shell to have negligible stress effect on the cavity block. The Al cans are gold coated on both sides to reduce radiative heat transfer.

**Molecular clock**

A clock stability of $\delta f/f < 5\times10^{-16}$ at orbit time is required to achieve the mission goal of the experiment. The clock instrument is based on optical frequency modulation (FM) spectroscopy techniques applied to a molecular species in the gas phase. The laser frequency is referenced to an optical resonance of the employed molecules – hyperfine transitions in Iodine ($I_2$) in this case. A laser can be locked to a molecular resonance by FM techniques similar to the PDH technique. The technique employs two overlapping counter-propagating beams passing through the gas sample for

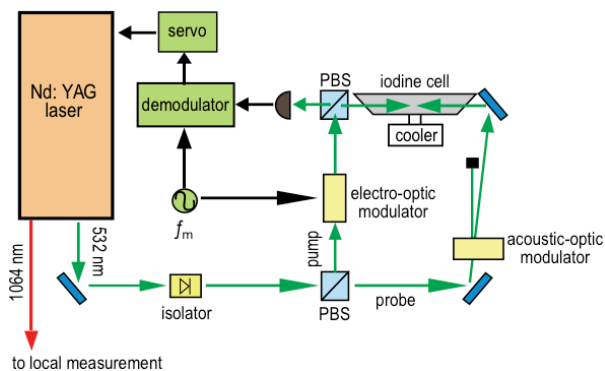

Fig. 6: Functional diagram of a laser stabilization scheme using the frequency modulation spectroscopy technique.

Doppler-free resolution of narrow transitions as shown in Figure 6.

Iodine clocks operating at 532 nm using this modulation transfer spectroscopy (MTS) technique have a long-term development advantage. A number of different setups allowing a simple implementation have been developed in the context of other space missions like LISA and many issues limiting the long-term performance have been identified[21]. In particular, MTS has been applied extensively to hyperfine transitions in molecular $I_2$ using frequency doubled Nd:YAG lasers. Nd:YAG lasers feature a superior intrinsic frequency and intensity stability, while $I_2$ features strong, narrow hyperfine transitions at 532 nm. These can be addressed with a frequency-doubled portion of the fundamental output of a Nd:YAG laser. The natural linewidth of the commonly used $a_{10}$ transition is 280 kHz. Taking into account pressure and power broadening the useable linewidth is typically on the order of 800 kHz.

To date, the best stability reported[22] at the 5800s orbit integration time is ~ $4\times10^{-15}$, within a factor of 8 of the requirement for the present mission. Based on the shot noise seen at 1 sec time scale it appears that an Iodine clock using the MTS technique is capable of meeting the requirement of the mission of a frequency stability of $5\times10^{-16}$ at orbit time. The presently observed degradation is likely to be due to the imperfect thermal control of the cavity used as a reference. Here we assume that this problem will not exist in the flight experiment due to the extreme thermal control and mechanical isolation planned for the cavities.

The optics design features fiber-coupled components that allow for modularity, reduced size, weight and power consumption. The setup will be built using a base-plate made of thermally and mechanically stable material such as Zerodur, SiC, or ULE. The optical components will be fixed to the base-plate either using hydroxide-catalysis bonding or adhesive patching using a space qualified two-component epoxy. Since mechanical instabilities in state-of-the-art setups are a major cause of the noise floor, a bonded setup is most likely to outperform the stability of state-of-the-art breadboard setups.

For detection of the MTS signal, a noise-canceling detector will be used. These provide an excellent signal-to-noise ratio in the presence of laser intensity noise. From the sensitivity of laboratory setups we can derive the environmental requirements to achieve a clock stability of $5\times10^{-16}$ at orbit time. These are summarized in Table 1. We believe they can be met by careful design and optimal choice of components.

Table 1: Environmental requirements for a MTS Iodine clock with a target stability of $5\times10^{-16}$ at orbit time.

| Parameter | Coefficient | Requirement (5800 s) |
|---|---|---|
| Temperature | 1 kHz/K | 280 µK |
| Laser intensity | 100Hz/mW | 2.8 µW |
| Dimensional | 10kHz/mrad | 28 nrad |
| Magnetic field | $10^{-15}$/Gauss | 0.01 Gauss |
| Pressure ($I_2$) | 3KHz/Pa | 1µPa |
| PDH offset voltage | 3 µV/Hz | 850 nV |

## Optics

A schematic of the optical layout for the instrument is shown in Figure 7. The signals are derived from a Nd:YAG laser frequency stabilized by locking the laser to a hyperfine transition of $I_2$. The stabilized laser light is transferred by fiber to a pair of identical optical trains, each of which couples to a Fabry-Perot resonator arranged in an orthgonally-crossed pair.

Each optical train frequency shifts and phase modulates the $I_2$-stabilized input light using an acousto-optic modulator (AOM) and an electro-optic modulator, allowing for the input laser light to be independently locked to the resonant frequency of each cavity using the PDH technique. The science signals then originate if a difference exists between the resonant frequency of each cavity and the clock laser frequency at orbital period (KT) or between the resonant frequencies of the cavities at roll period (MM). These signals appear in the frequency of the RF drive to the AOMs for each cavity.

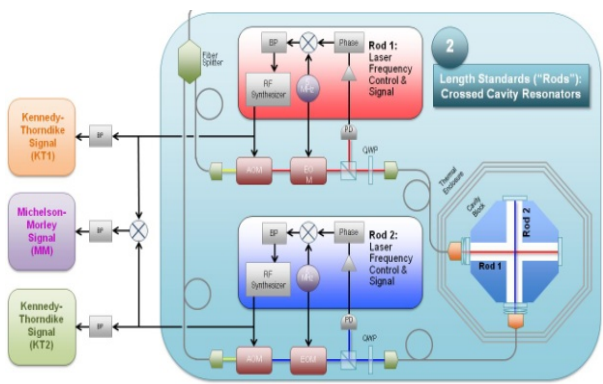

Fig. 7: Optical setup showing a pair of cavities referenced to the clock input signal.

## Electronics

The payload electronics needs to provide the usual sensor functionality as well as specialized circuits to handle the high precision optical and thermal subsystems of the payload. Key circuitry for the science readout is that used to lock the laser frequency to $I_2$ and to the resonant frequencies of the cavities. With the very high measurement precision required, the critical components in these circuits will need to be carefully selected to provide the needed stability and fidelity of signals. Thermal control of some components will also be necessary. The thermal sensing and control scheme for the length reference needs to use very stable components and balanced electronics to reach the needed thermal stability for the sensor. Thermal stability is achieved through a combination of selectable measurement ranges, high resolution (24-bit) sensing, bridge electronics and temperature-controlled critical components in the feedback circuitry. In addition, the very high resolution sensing that is used for feedback control of the heaters makes use of a processor for implementation of the control laws. Nevertheless, all component requirements appear to be within the current state-of-the-art.

## Instrument Error Budget

The top-level instrument random error budget for the KT experiment is shown in Table 2. The total experiment error for the KT measurements is the root-sum-square of the noise levels of the individual components contributing to the budget. The KT measurement period is 5,800 sec (orbit period) while the MM measurement period is nominally 250 sec (1/2 spacecraft spin period) resulting in more relaxed requirements. After two years of integration the accuracy of both science measurements meets the goal of $\delta c/c$ $10^{-17}$ with margin. The systematic noise is estimated as $< 3 \times 10^{-18}$.

Table 2: KT random error budget.

| Source | Requirement | Estimated value |
| --- | --- | --- |
| Cavity temp. stability | $10^{-8}$ K pp | $3 \times 10^{-9}$ K pp |
| Mirror noise | $5 \times 10^{-16}$ $\delta l/l$ | $3 \times 10^{-16}$ $\delta l/l$ |
| Cavity gas press. | $10^{-4}$ torr | $10^{-5}$ torr |
| Shot noise | $1 \times 10^{-16}$ $\delta l/l$ | $2 \times 10^{-18}$ $\delta l/l$ |
| Satellite pointing | $1 \times 10^{-17}$ $\delta l/l$ | $1 \times 10^{-17}$ $\delta l/l$ |
| Molecular clock | $5 \times 10^{-16}$ $\delta f/f$ | $3 \times 10^{-16}$ $\delta f/f$ |
| Comparator | $2 \times 10^{-16}$ $\delta f/f$ | $2 \times 10^{-16}$ $\delta f/f$ |
| RSS noise | $1 \times 10^{-15}$ $\delta c/c$ | $5 \times 10^{-16}$ $\delta c/c$ |
| 2 yr integration with 2 systems | $1 \times 10^{-17}$ $\delta c/c$ | $3 \times 10^{-18}$ $\delta c/c$ |

## SPACECRAFT

In this section we describe those aspects of the spacecraft design that are important for extracting the science results. The primary functions of the spacecraft are to provide power and a layer of thermal control to the instrument, maintain sun pointing and roll, and handle communications. The functions that have the largest effect on the science are thermal, pointing and roll.

**Thermal**

As described above, a sophisticated thermal enclosure is needed for the instrument to filter temperature variations at the orbital period. The spacecraft is used to provide a first layer of isolation from external couplings to heat sources and sinks. For the optimum Sun-synchronous orbit there appears to be a wide margin of safety using the 6-layer instrument thermal control system. For an equatorial orbit (which we treat as near to worst case), much larger temperature swings exist from light to shadow, but even these are easily filtered with a passive system. Adding two layers of active thermal control would give wide margins once again.

**Pointing**

The spacecraft pointing requirement is driven mostly by the desire to maintain a stable thermal profile while in sunlight, and to minimize angular accelerations on the optical cavities. These accelerations affect the KT measurement to the extent that they induce a length change in the cavity at orbital period. The constraint here is very weak, with an excursion of a radian at orbital period being allowable. Thus solar radiation effects are the dominant issue. The pointing constraint is set at 0.5 deg to avoid sunlight falling on the side panels of the vehicle.

Another issue is the differential red shift between the cavities and the Iodine cells due to the offset of their 'centers of frequency' along the roll axis and its projection onto the local gravity vector. We plan to equip the spacecraft with a pair of star trackers to obtain precision pointing information relative to inertial space. With a pointing knowledge of ± 10 arc-sec it appears possible to correct for this effect to a negligible level. We also note that the amplitude of the effect is cyclical with a 1-year period while the KT signal is fixed in inertial space, allowing separation even without correction.

**Roll**

Spacecraft roll is most important for the MM experiment because it provides the mechanism for measuring angle dependent effects. The highest sensitivity is obtained by having both cavities aligned normal to the roll axis. This is also advantageous because roll rate perturbations become common mode relaxing the mechanical stability requirement. For KT, roll is not needed and the main requirement is on angular rate variations that might have components near the orbital period. Fortunately there is no significant mechanism for generating such effects. A bound on the allowable rate can be obtained by considering centrifugal forces acing on a cavity. For a roll rate of $2 \times 10^{-3}$ Hz the limit on variability is $\sim 0.5\%$. This is easily measured with a simple star tracker with stellar determinations to $\sim 100$ arc sec. Thus even if a roll variation exists it will be relatively easy to remove it from the KT data.

Roll also introduces significant differential red shifts between the cavities and clocks, and gravity gradient effects in the cavities. These appear at different frequencies from the MM and KT signals because they are aligned with the local gravity vector, not with inertial space. If necessary they can also be modeled using the spacecraft attitude information. Roll also serves as an attenuator of thermal effects from variable Earthshine to negligible levels as determined by finite element modeling.

**Orbit**

The mission orbit altitude is derived from several considerations. First, the sensitivity of the KT experiment decreases slowly with orbit altitude. Second, the cavity deformation due to atmospheric drag decreases rapidly with altitude. On the other hand, the low altitude minimizes the need for radiation shielding and hardened components. Also, a low altitude allows the spacecraft to de-orbit naturally within the 25-year limit set by NASA, thus obviating the need for a costly de-orbit system. A 650 Km altitude has been selected as a good compromise between these competing factors, with drag effects on science being negligible. An approximately circular orbit is preferable because it will provide the highest differences in velocity vector.

A sun-synchronous orbit will maintain approximately a fixed orientation relative to the sun throughout the active lifetime of the mission, enhancing the thermal stability of the science payload. To minimize eclipse time (which occur over a few weeks each year) and drag variations, a 6 AM dawn-dusk sun-synchronous orbit is the mission baseline. We also note that the plane of the sun-synchronous orbit passes through 11.2 h and 23.2 h of Right Ascension twice per year, giving maximum sensitivity to the direction of the Earth's motion relative to the CMB.

## DATA ANALYSIS

The basic data analysis approach is a weighted least squares fit[23] of the time-domain models of the various signatures of the MM, KT, and Lorentz coefficients, and other perturbations as needed. For the MM measurement, the time signature is derived from the time-history of the inertial orientation of the instrument (three Euler angles, or three dimensional unit vector). For the KT measurement the time signature is derived from the three-dimensional inertial spacecraft velocity time-history. The models are linear with respect to the parameters to be estimated, leading to a straightforward data analysis.

The science measurements are made in a reference frame fixed to the spacecraft instrument, whose orientation and position is changing with respect to the inertial reference frame. However, the science analysis must be performed in an inertial reference frame, preferably aligned with the CMB. Therefore, an inertial CMB frame is defined, using the direction of the Sun's motion relative to the CMB (declination = –6.4 deg, and RA = 11.2 h). Secondary reference frames are then generated to couple this frame to the orientations of the length references. These frames are Sun-centered (inertial), Earth-centered (non-inertial) and spacecraft-fixed (non-inertial) coordinate systems. Transformations between the various frames are then derived, allowing instrument pointing and spacecraft inertial velocity information to be transformed from the spacecraft or Earth-centered frames to the inertial frame. The data will be segmented into individual orbits for analysis of velocity effects, and into integral numbers of roll cycles for directional effects, as seen in inertial space.

The weighted least squares fit will produce optimal estimates of the parameters describing the ellipsoids by minimizing the $\chi^2$ of the post-fit residuals. The weighted least squares approach is ideally suited to handle data with dropouts, sections with excess noise, as well as low frequency (1/$f$) noise. The large volume of data collected by the instrument means that some outliers may occur statistically. Any outlier that is inconsistent with the statistical expectation will be removed prior to estimation of the coefficients. The validity of the coefficient estimates will be demonstrated by comparing the post-fit residuals to independent estimates of the intrinsic instrument noise ($\chi^2$), and by checking the whiteness of the residuals.

### Kennedy-Thorndike Coefficient

The KT coefficient is obtained from a data analysis using eq. 1 to look for a preferred frame effect manifested by the velocity dependence of $\delta c/c$ independent of orientation. From the form of eq. 1 this signal will be at orbital period. The derived quantity is then the $\theta$-independent term in eq. 1: $\varepsilon_{KT}(v/c)^2$. Since it is very difficult to measure $c$ directly with high precision, the only way this analysis can generate an observable $\delta c$ is if $v$ varies. To obtain extra sensitivity over ground we make use of the orbital velocity of the spacecraft to maximize the change in $v$, denoted $\delta v$. By convention $v$ is taken as the velocity relative to the CMB and any signal detected is referenced to this frame. To get the sensitivity to the KT coefficient we then multiply the $\delta c/c$ data by $c^2/v\delta v$. In a circular orbit at 650 km altitude we obtain $2.2 \times 10^7$ for this quantity giving a high sensitivity to small effects. The resulting limit on $\varepsilon_{KT}$ is $< 4 \times 10^{-10}$. This can be compared with the ground limit of $\varepsilon_{KT} < 4 \times 10^{-8}$ supporting our claim of 100x improvement in the resolution of this parameter. Testing for other potential preferred frames would involve a similar analysis aligned with other directions of interest.

### Coefficients of Lorentz Violation

The SME provides a more general formalism than RMS for interpreting the data. The coefficients of Lorentz violation represent the most general set of Lorentz violations that can occur within the SME. To obtain these coefficients the $\delta c/c$ data is re-analyzed to derive first and second harmonic information in two orthogonal planes. The fitting equation is of the form:

$$\delta c/c = A\sin\Phi + B\cos\Phi + C\sin2\Phi + D\cos2\Phi \quad (2)$$

where the amplitudes A, B, C, and D contain linear combinations of the coefficients of Lorentz violation and $\Phi$ is a phase angle relative to inertial coordinates. The coefficients of Lorentz violation measured by the experiment will be for mixed photon and electron sectors. They are relevant as the first order, velocity-independent terms in the SME as applied to cavities[24]. These are important results for theorists attempting to unify the SM with gravity because they are linked directly to the interaction fields currently assumed to be Lorentz invariant. These terms will be measured with a resolution of ~ $3 \times 10^{-17}$. Velocity-dependent terms also enter eq. 2 and are purely from the photon sector. These are bounded at the $10^{-13}$ level, a factor of ~15 better than ground measurements. In this case $\Phi$ is a phase angle representing the direction of the velocity vector.

As mentioned earlier, in the case of the KT-style measurements the SME coefficients are not yet quantified but are expected to be analogous to the terms described above. Theoretical work to define these parameters in the SME and their impact on the underlying physics is in progress elsewhere[25].

**Michelson-Morley Coefficient**
The MM coefficient is obtained by comparing the apparent lengths of two rods perpendicular to each other using light beams as the yardsticks. The data is the $\delta c/c$ signal as a function of orientation, but now one assumes equation 1 is correct and computes the coefficient of the $\sin^2(\theta)$ term: $\varepsilon_{MM}(v/c)^2$. The MM coefficient $\varepsilon_{MM}$ is larger by a factor $(c/v)^2$ relative to the basic measurement, $\delta c/c$, to give a parameter that can be compared directly with other experiments. As mentioned above, by convention $v$ is taken as the velocity relative to the CMB and any signal detected is referenced to this frame. We expect to resolve $\varepsilon_{MM}$ to ~ $1 \times 10^{-11}$. Alternatively, one can search for a $\sin^2(\theta)$ dependence along any axis, in which case the total amplitude of the MM term would be constrained. This would amount to a generalized search for a preferred frame in any chosen direction.

While it is unlikely that the MM measurement will be substantially better than on the ground, it does serve as a useful diagnostic for the anisotropy measurements because it provides a cross check on the behavior of the 'rod'. Also, as mentioned earlier, there has been speculation that LI violations might be different in space because of the near absence of matter surrounding the apparatus.

**Systematic effects:**
Systematic effects such as instrumental frequency drifts during a roll cycle, frequency offsets due to spin speed changes, gravity gradient, thermal and centripetal acceleration signatures will need to be included in the model as necessary. First order models for each of these effects have already been derived and used to supplement the instrument error budget.

Many of the systematic effects are naturally separated from the MM and KT effects when the instrument is operated for longer than 1 year. This is because solar and terrestrial effects (thermal, gravity gradient, redshift) will vary throughout the year as the position and orientation of the spacecraft change relative to the Sun and the Earth, while LIV effects are fixed in the inertial frame. Consequently, these systematic effects are averaged-out over the course of a year, which drives the requirement for a 2-year mission to allow for checks in this area. In the frequency domain the majority of systematic effects occur at the spacecraft spin and orbit frequencies, while the MM violation occurs at the frequency 2×spin ± 2×annual and the KT violation occurs at orbit ± annual.

In addition, a set of on-orbit tests are planned to calibrate the models for each systematic effect. For example, spin related effects will be calibrated by operating the instrument at several different spin rates. Lorentz violations are simple functions of orientation, while thermal effects are nonlinear due to the associated thermal time constant and mechanical distortion due to centripetal acceleration is proportional to the square of the spin rate. Therefore, the spin rate calibrations enable many systematic effects to be separated from the Lorentz Violation signals. This calibration procedure will set the final spin rate for the science data collection phase of the mission.

## FUTURE PROSPECTS
The experiment described here makes use of well-tested technology that is ready to transition to space use with minimal further development. It is also of interest to consider what other advances might be made on a somewhat longer time scale. It can be seen from Table 2 that the limiting factors for the present design are the performance of the iodine molecular clock and cavity mirror noise. Hall has pointed out that other molecular species could be chosen, such as $C_2H_2$ or CO. These compounds have a much lower absorption cross-section than $I_2$, but their spectra have much better separated lines. This reduces unwanted fluorescence that can increase the noise in a Doppler free measurement. With the use of a resonant cavity it is possible to enhance the absorption, greatly improving the S/N[26]. It appears possible that CO for example, could reach $\delta f/f \sim 3 \times 10^{-17}$ at orbital period. In addition laser-cooled single atom clocks can have extremely low statistical noise and work is in progress to build a space qualified Yb or Sr clock[27]. These systems would then essentially drop out of the error budget, putting the onus on cavity noise. Here we note that ultra-low-loss coatings are being developed for the LIGO experiment and a factor of 2 noise reduction seems possible. This combined with four 20 cm cavities and a 4-year mission lifetime could push the effective cavity noise at orbital period to $\sim 10^{-16}$ and the resulting mission resolution to $\sim 8 \times 10^{-19}$.

Another possibility is to make use of the International Space Station (ISS). While this platform has a significant level of vibration noise, it allows payloads of a few hundred kg and provides ample power to operate. Perhaps its most unique capability for the present experiment is the possibility of cryogenic operation using cryocooler technology. This would allow a significant improvement in cavity stability and lower the thermal noise that currently limits frequency discrimination. Some steps are already being taken in this direction with the development of the Space Optical Clock payload, planned for deployment on the Columbus module of the ISS[27].

## CONCLUSION

We have described the essential features of a small flight experiment that is capable of significantly improving the experimental underpinnings of special relativity. At least a factor of 100 gain in sensitivity is expected over current measurements in an analog of the KT experiment. With a more advanced unit a factor of 1000 gain seems possible. Measurements of the MM type would also be made with a sensitivity comparable to ground. Parameters in the SME could be derived with some limits at least a factor of 15 better than ground.

## ACKNOWLEDGEMENTS

The flight experiment described here is a collaboration of institutions in the USA, Saudi-Arabia and Germany. The authors thank John Hall from NIST, Peter Worden, Belgacem Jaroux and Brian Lewis from NASA Ames and Turki Al-Saud and Haithem Altwaijry from KACST for their support. We also thank Claus Braxmaier, Achim Peters, Claus Laemmerzahl, Hansjoerg Dittus, Klaus Doringshoff and other members of the German team for their parallel work developing a flight qualified molecular iodine clock, and Jeff Scargle of Ames and Ke-Xun Sun of U. Nevada for technical support.